\documentclass[aip,reprint,floatfix,superscriptaddress,showpacs,amsfonts,amssymb,amsmath,preprintnumbers]{revtex4-1}

\usepackage{graphicx}
\usepackage{float, amsmath,verbatim,amssymb,stmaryrd}
\usepackage{epsfig}
\usepackage{xcolor}
\usepackage{comment}

\usepackage{xcolor}
\newcommand{\sm}[1]{{\color{blue}#1}}

\begin{document}


\title{Are we ready to transfer optical light to gamma-rays?}

\author{M. Vranic}
\email{marija.vranic@ist.utl.pt}
\affiliation{GoLP/Instituto de Plasmas e Fus\~ao Nuclear, Instituto Superior T\'ecnico-Universidade de Lisboa, 1049-001, Lisbon, Portugal}
\author{T. Grismayer }
\affiliation{GoLP/Instituto de Plasmas e Fus\~ao Nuclear, Instituto Superior T\'ecnico-Universidade de Lisboa, 1049-001, Lisbon, Portugal}
\author{S. Meuren }
\affiliation{Department of Astrophysical Sciences, Princeton
University, Princeton, New Jersey 08544, USA}
\author{R. A. Fonseca}
\affiliation{GoLP/Instituto de Plasmas e Fus\~ao Nuclear, Instituto Superior T\'ecnico-Universidade de Lisboa, 1049-001, Lisbon, Portugal}
\affiliation{DCTI/ISCTE Instituto Universit\'{a}rio de Lisboa, 1649-026 Lisboa, Portugal}
\author{L. O. Silva}
\email{luis.silva@ist.utl.pt}
\affiliation{GoLP/Instituto de Plasmas e Fus\~ao Nuclear, Instituto Superior T\'ecnico-Universidade de Lisboa, 1049-001, Lisbon, Portugal}

\date{\today}

\begin{abstract}

Scattering relativistic electrons with optical lasers can result in a significant frequency upshift for the photons, potentially producing $\gamma$-rays. 
This is what linear Compton scattering taught us. Ultra-intense lasers offer nowadays a new paradigm where multi-photon absorption effects come into play. 
These effects can result in higher harmonics, higher yields and also electron-positron pairs. 
This article intends to discriminate the different laser scenarios that have been proposed over the past years as well as to give scaling laws for future experiments. 
The energy conversion from laser or particles to high-frequency photons is addressed for both the well-known counter propagating electron beam-laser interaction and for Quantum-electrodynamics cascades triggered by various lasers. 
Constructing bright and energetic gamma-ray sources in controlled conditions is within an ace of seeing the light of day. 


\end{abstract}



\maketitle

\section{Introduction}

Ultra-intense laser-matter interactions can generate $\gamma$-rays. 
Several recent experiments have obtained hard X-rays in the laboratory using counter-propagating electron beams and laser pulses  \cite{MalkaNature, Sarri_compton_exp, Chen_compton_exp, Powers_compton_exp, Khrennikov_compton_exp}. 
A review on laser-wakefield acceleration (LWFA) - based light sources summarizes other viable configurations for generating energetic photons and the properties of radiation that could be obtained with a reflection to prospective applications \cite{Review_Albert_Thomas}. 

Multiphoton Thompson scattering was observed in a recent experiment by Yan et al \cite{Umstadter_multiphoton_t}. 
The next goal is to obtain a high conversion efficiency of the electron energy into high-frequency radiation.
This requires operating the source in a regime of significant radiation recoil, in classical or quantum regime of interaction. 
First evidence of electron slowdown in a laser field \cite{cole2017_experimentalRR, poder2017_experimentalRR} are consistent with the classical radiation reaction predictions for scattering a LWFA electron bunch and a laser pulse \cite{First_PRL}.  
However, there are still many open questions regarding the transition from the classical to the quantum radiation reaction dominated regime, that has generated keen interest in the last few years \cite{Ridgers_quantumRRnew, Piazza_qed_energyspread, Piazza_QRR_FD, Dinu_harvey, Yoffe_evolution, Vranic_quantumRR, Ridgers_RR_spinetto, niel2017quantum}. 

The radiation reaction dominated regime (i.e. the regime where
	particles lose a substantial amount of energy through radiation
	emission) can be reached either by using higher laser intensities or
	more energetic electrons. The particles in this regime either emit a
	few very energetic photons (quantum radiation reaction) or a large
	number of low-energy photons successively (classical radiation
	reaction) \cite{RevModPhysdipiazza}.
Current published energy record for LWFA electrons is 4 GeV and those electrons are obtained using a 16 J laser \cite{Leemans_4gev}. 
In the next generation laser facilities, electron energies on the order of 10 GeV are expected. 
Pairing such electron beams with intense lasers can provide access to more extreme regimes of interaction. 
It may even be possible to obtain multiple electron-positron pairs from electron beam-laser collision. 
In the two-step approximation a particle first emits a high-energy photon, which then decays into an electron-positron pair via the Breit-Wheeler (BW) process in an intense electromagnetic background.
A milestone experiment was performed at SLAC \cite{E144slac1}, where BW pairs were produced in a collision of a 46.6 GeV electron beam from a linear accelerator with a green laser at the intensity of $\sim 10^{18}~\mathrm{W/cm^2}$. 
The next generation of laser facilities is expected to deliver a much higher pair yield \cite{Lobet_Davoine_pairs, Zhu_dense} even using electrons with lower energies (on the order of a few GeV). 
Further laser-electron scattering experiments are planned, both relying on LWFA and conventional accelerators to provide the electrons.

Another way to generate $\gamma$-rays and pairs with intense lasers are QED cascades \cite{Bell_Kirk_2lasers, RevModPhysdipiazza}. 
One of the most popular configurations is to use two colliding lasers to create an intense standing wave \cite{Zhidkov_RR_laser, model1bell, Bulanov2006, Nerush_laserlimit, Thomas_POP_2016, Mironov_oblique} whose nonlinear evolution in the presence of self-generated plasma can be studied theoretically resorting to QED-PIC simulations \cite{Gremillet, lobet, Vranic_merging, Elkina_rot, Ridgers_solid, Nerush_laserlimit, Bell_Kirk_MC, Gonoskov2015review, Basmakhov}. 
New laser facilities will access extreme regimes of interaction \cite{ELI_WhiteBook, QED_at_ELI, ELI_NP_2016, WEBER_P3} where we can expect an abundance of electron-positron pairs. This motivated a lot of theoretical effort to improve our understanding  of QED cascades, considering configurations using multiple lasers \cite{Gelfer, MV_ppcf2016, Gonoskov_PRX}, as well as the challenges concerning the cascade seeding \cite{Jirka, ThomasQED}.
Among the proposed solutions for the seeding problem, it was proposed to use solid targets \cite{Kostyukov_new, jirka2017_target}. Configurations with nanowires were also proposed, to enhance the laser heating of the target electrons \cite{Bertrand_Nanowires_2018}. 

In this manuscript, we focus on the gamma-ray emission. 
We revisit the two main configurations: laser-electron beam scattering and the two-laser QED cascade. 
The first aim is to connect the hard-photon emission in both of these scenarios and bring an intuitive understanding of why even in a QED cascade, the classical absorption is more important than the instantaneous quantum absorption.
The ratio of the classical and quantum absorption has been found previously using QED calculations \cite{meuren_article} to be proportional to the square of the local normalized vector potential $a_0$, where $a_0=0.86 \sqrt{I[10^{18}~\mathrm{W/cm^2}]} \lambda[\mu \mathrm{m}]$ for linearly polarized lasers and $a_0= 0.61 \sqrt{I[10^{18}~\mathrm{W/cm^2}]} \lambda[\mu \mathrm{m}]$ for circular polarization. 
In other words, this ratio is proportional to the laser intensity, and for high laser intensities, the classical absorption dominates. 
Here, we show how the electron-photon scattering in a QED cascade can be mapped to a simple laser-electron scattering scenario for which the quantum vs. classical absorption ratio has already been calculated. 
The second aim is to estimate how much energy is converted to hard photons in both configurations and identify the relevant regimes of interaction. 
We also give a brief summary of the scaling laws for evolution of the electron energy distribution function in the laser - electron beam scattering. This is of importance for planning of experiments, because the electron beam properties imprint on the emitted radiation. 
The ideas and scalings presented in this manuscript are relevant for the multi-petawatt laser projects such as ELI \cite{ELI_WhiteBook}, Apollon \cite{Apollon} and CoReLS \cite{Corels} that aim to reach unprescedented laser intensities, as well as FACET-II \cite{FACET} and LUXE \cite{Luxe, Luxe_exp_theory} that plan to perform laser scattering experiments using 10 GeV-class high-quality electron beams.
The manuscript is organized as follows. 
In the following section, we discuss the ratio of classical vs. quantum absorption in a scattering of a single electron with one wave.
We mostly discuss the counter-propagating geometry, because it provides the highest energy photons. 
We then extend the ideas presented for a single wave-electron interaction, to a lepton interacting with a standing wave formed by two colliding lasers. 
We then review the scaling laws and the energy conversion expected for the electron-laser scattering configuration. 
We finally discuss the radiative absorption in a two-laser cascade. 
We distinguish two regimes: a regime of controlled radiation emission, when the wave is not severely affected by the presence of the plasma and a regime where the plasma density is high enough to disrupt the wave (this can happen because the target is dense in the beginning, or due to the considerable production of electron-positron pairs).
We identify the parameters where the pair production is low enough to operate  a controlled $\gamma$-ray source in a low-absorption regime with solid hydrogen targets already available in the laboratory. 
The results are supported with QED-PIC simulations for a range of parameters, both in 2D and 3D geometry.

\section{Quantum laser absorption vs. classical laser absorption}

Quantum vs. classical laser absorption by a single electron was first considered by Meuren et al. \cite{meuren_thesis, meuren_article} using the QED formalism. Here, we provide an intiutive picture for the corresponding scaling laws. 

\subsection{Laser-electron scattering}

\begin{figure*}
	\includegraphics[width=0.8\textwidth]{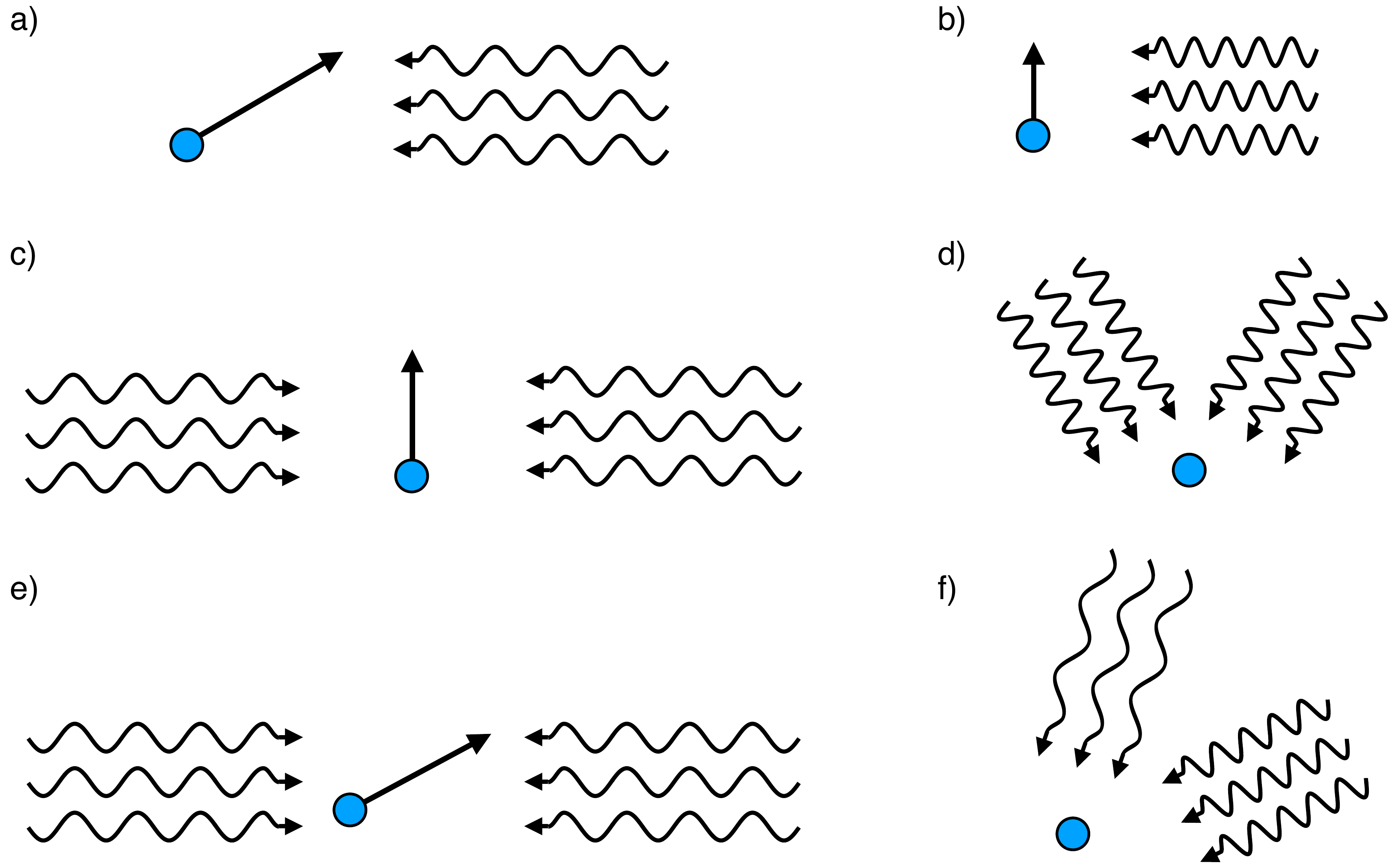}
	\caption{Geometry of the interaction in laboratory (left-hand side) and boosted frames (right-hand side). a), b) Laser-electron head-on configuration; c), d) Two-laser standing wave, when interacting particle instantaneous momentum is perpendicular to the laser axis; e),f) Two-laser standing wave interacting with a particle at an oblique angle.  }
	\label{compton_sketch}
\end{figure*}
	
The first studies dealing with interaction of electrons with intense laser beams date back to a few years after the invention of the laser \cite{Lowell&Kibble, Kibble}. It was found that a mass change induced in the electron (considered initially at rest) by the external field of the laser shifts the wavelength of the scattered photons by an amount depending on the intensity of the incident beam. Furthermore, the absorption of multiple laser photons becomes possible, which facilitates the emission of harmonics. If the electron is initially relativistic, the scattering with the laser photons can result in the emission of $\gamma$-rays. 

Effectively, only a finite space-time region is relevant for the photon emission process. It is characterized by the so-called formation length $l_f$ that depends on the wave intensity \cite{Ritus_thesis}
\begin{equation}\label{form_length}
\frac{l_f}{\lambda}\sim\frac{1}{a_0},    
\end{equation}
where $\lambda$ is the wavelength of the wave. 
Equation (\ref{form_length}) is valid for $\chi_e\lesssim1$, where the definition of quantum parameter for electrons counter-propagating with an optical laser ($\lambda=1~\mu$m) can be approximated as $\chi_e\simeq 5\times10^{-6}\gamma_0 a_0$.

According to Eq.\,(\ref{form_length}) the radiation is formed over multiple laser cycles if $a_0\ll1$. In this regime it is useful to introduce the so called "dressed momentum" for the emitting electron, which is obtained by averaging the classical canonical momentum over a laser period \cite{Ritus_thesis, QEDbook, Lowell&Kibble}. Correspondingly, the classical mass change mentioned above appears.

For $a_0\gg1$ we have $l_f\ll\lambda$ and the radiation process happens almost instantaneously in an effectively constant background field. The concept of dressed momentum in this regime becomes meaningless \cite{yakimenko2018prospect, meuren_thesis,meuren_article}. Instead, a semiclassical approximation becomes possible, i.e. the radiating electron follows a classical trajectory between localized emission events \cite{baier_Bfield, Antonino_nonlinearC}.

In this regime we can distinguish two types of energy transfers: classical and quantum absorption \cite{meuren_thesis, meuren_article}. The term quantum laser absorption $\epsilon_{LQ}$ is defined as the direct contribution of the laser photons that scatter with the electrons during the quasi-instantaneous emission process. 
Any laser absorption that occurs elsewhere (the laser energy invested in the electron acceleration) can be described classically, and is incorporated into the instantaneous electron energy contribution $\epsilon_e$.
The analysis of the energy transfer from electrons to photons allows estimating the importance of quantum laser absorption during the  electron interaction with an intense laser beam.


The scattering process is affected by the presence of the intense laser field in two ways. 
One aspect is that there is a temporary momentum transfer from the laser to the electrons: the laser induces electron oscillations. 
A relativistic electron counter-propagating towards an intense laser with normalized vector potential $a_0$, acquires an additional transverse momentum on the order of $\sim a_0 mc$ during the interaction, while the effect on the longitudinal momentum is small if $a_0\ll \gamma_0$. 
For circular polarization, the additional electron transverse momentum is constant in magnitude but periodically changes the direction revolving around the propagation axis, while for linear polarization, the direction is fixed, but the magnitude changes periodically. 
The second effect of the very intense field to the scattering process is that the local field energy density is high, which means there can be many photons within the interaction volume. 
Such photon density allows to have frequent repeated scatterings, as well as to absorb more than one photon at a time in a single scattering.

The following paragraph illustrates intuitively, by comparing the typical timescales of acceleration and emission, how the factor $a_0^2$ between the classical and quantum photon absorption might arise. 
Let us assume that $a_0\gg 1$ and the particle emits a hard photon not more than once during one full laser cycle. 
The photon formation time is $\tau_f=l_f/c$, where $c$ is the speed of light. 
The laser period is $T_L$.
As $T_L/\tau_f\sim a_0$, the total work of the laser electric field during one period on one particle is $a_0$ times larger than during one emission event. 
However, at relativistic intensities, an electron can take much more energy from the wave, on the order of $\xi\sim a_0^2 mc^2$ (assuming the electron is initially at rest).
This is because a relativistic electron can have a longer effective interaction time, and one oscillation can last longer than a simple laser period $T_L$  
(for example, if an electron co-propagating with the wave).
From there we get that $T_\mathrm{eff}\sim a_0T_L$ and $T_\mathrm{eff}/\tau_f\sim a_0^2$. 
One should note that the value of $T_\mathrm{eff}$ depends on the initial particle energy and the scattering angle, but it is never smaller than $T_L/2$.
If we have $n$ emissions during one laser cycle, then the average time between two emissions is $T_\mathrm{eff}/n$. 
The relevant ratio then becomes $T_\mathrm{eff}/(n \tau_f)\sim a_0^2/n$.
If a particle is counter-propagating with a wave, then most of the energy of the emitted photons is invested by the electron. 
However, the laser has to invest a few photons (even if not many) into the electron acceleration (classically), and also during the actual emission event in order to facilitate it (quantum absorption). One can show that the classical vs. quantum absorption of the laser photons scales as $a_0^2$. 
As the quantum absorption is negligible for $a_0\gg1$, it is not considered in the QED-PIC algorithms. 
However, the classical absorption that happens due to the laser interaction with the plasma particles (i.e. particle acceleration) is intrinsically included in the PIC algorithm.

\subsection{Mapping photon emission in a QED cascade produced by two colliding lasers with the photon emission in laser-electron scattering}

The ideas presented for a laser - electron scattering do not trivially port to the standing wave configurations with multiple lasers. 
Particles can be initially at rest. 
The particles first get accelerated, then lose energy due to radiation emission, and then get re-accelerated\cite{Fedotov_cascade} by the electric field in the standing wave. 
All the energy radiated to high-frequencies comes from the laser field: either through accelerating the electrons (classical absorption) or by providing photons for scattering to occur (quantum absorption). 
The question is: which channel is dominant?

We can make use of what we know about particle dynamics in a standing wave to establish a connection between this setup and the simple laser-electron scattering discussed in the previous section. 
Electrons (and positrons) gain momentum on the order of $a_0$ that is perpendicular to the laser propagation axis \cite{Fedotov_cascade,Elkina_rot,ThomasQED}. 
In the case of circular polarization, leptons keep rotating, always remaining perpendicular to the laser propagation. 
This allows for a simplified consideration, as the laser-photon scattering always occurs at the right angle. 

If the particle scatters with photons from only one laser, the situation is exactly the same as a scattering with one laser at 90 degrees. 
However, in principle, we could have a linear combination of $m$ photons from one wave and $n$ photons for the other participating in the scattering (not every combination is necessarily allowed, but here we are assuming a most general case).
As the particle Lorentz factor is typically on the order of $a_0$, the energy of an individual photon in the particle rest frame is $\xi'_{ph}\sim a_0~ \xi_{ph}$ regardless which laser the photon belongs to.
 Furthermore, the photons of both waves are practically co-propagating in the electron rest frame, as the Lorentz boost gave them a momentum in the same direction. 
 In other words, the electron sees them almost as one wave, apart from the tiny difference in the momentum ($\sim$ 1 eV for optical photons) perpendicular to the Boost direction (see Fig. \ref{compton_sketch} d)). 

Particle dynamics is slightly different when we have linearly polarized laser pulses forming the standing wave. 
The electric field is perpendicular to the laser propagation axis, but the magnetic field then rotates the momentum, and the particles can be found counter-propagating with one of the waves. 
In the electron rest frame, counter-propagating photons are upshifted, while the photons of the co-propagating laser are downshifted. The most energetic photons in the electron rest frame are the counter-propagating ones.  
We can then see the analogy with the laser-electron scattering emerge naturally, as one wave becomes more important then the other.
We note here that the normalized vector potential $a_0$ is Lorentz-invariant, which means that the wave with higher energy individual photons in the electron instantaneous rest frame has a higher energy density.  
There is a range of possible angles of incidence, with two limiting cases: particles being perpendicular to the laser propagation axis or counter-propagating with one of the lasers (see Fig. \ref{compton_sketch} e)).
In general, any of these configurations can be mapped to an electron-laser scattering
 
We have not discussed a decay of a hard photon into a pair here, because the number of emission events in a QED cascade surpasses the number of pair production events by orders of magnitude. 
They contribute little to the overall energy balance. 
Nevertheless, it was shown in Ref. \cite{meuren_article} that the ratio between the classical and quantum laser absorption for pair production is also $\epsilon_{CL}/\epsilon_{LQ}\sim a_0^2$.

\section{Scaling laws for laser-electron scattering}
\label{sec:dev_cas}

In a collision between an intense laser and an electron beam, the final electron energy after the interaction can be estimated as \cite{First_PRL}
\begin{equation}\label{efinal}
\gamma_F\simeq\frac{\gamma_0}{1+k \gamma_0}
\end{equation}
for $k<1$. The coefficient $k$ depends on the laser duration $\tau_0$ at FWHM and peak intensity $I_0$ in the following way
\begin{equation}\label{k_efinal}
k=3.2\times 10^{-5} ~I_{22}~\tau_0[\mathrm{fs}]~(1-\cos \theta)^2 \sm{,}
\end{equation}
where $I_{22}= I_0\left[10^{22}~\mathrm{W/cm^2}\right]$ and $\theta$ represents the angle of interaction (for counter-propagation the $\cos \theta\simeq-1$).
The counter-propagating configuration, therefore, corresponds to the strongest radiation reaction or losing a largest fraction of the electron energy. 
The energy converted to photons is then approximately
\begin{equation}
    \xi_{rad}[mc^2]\simeq \frac{k\gamma_0^2}{1+k\gamma_0} \sm{.}
\end{equation}

\begin{figure}
	\includegraphics[width=0.45\textwidth]{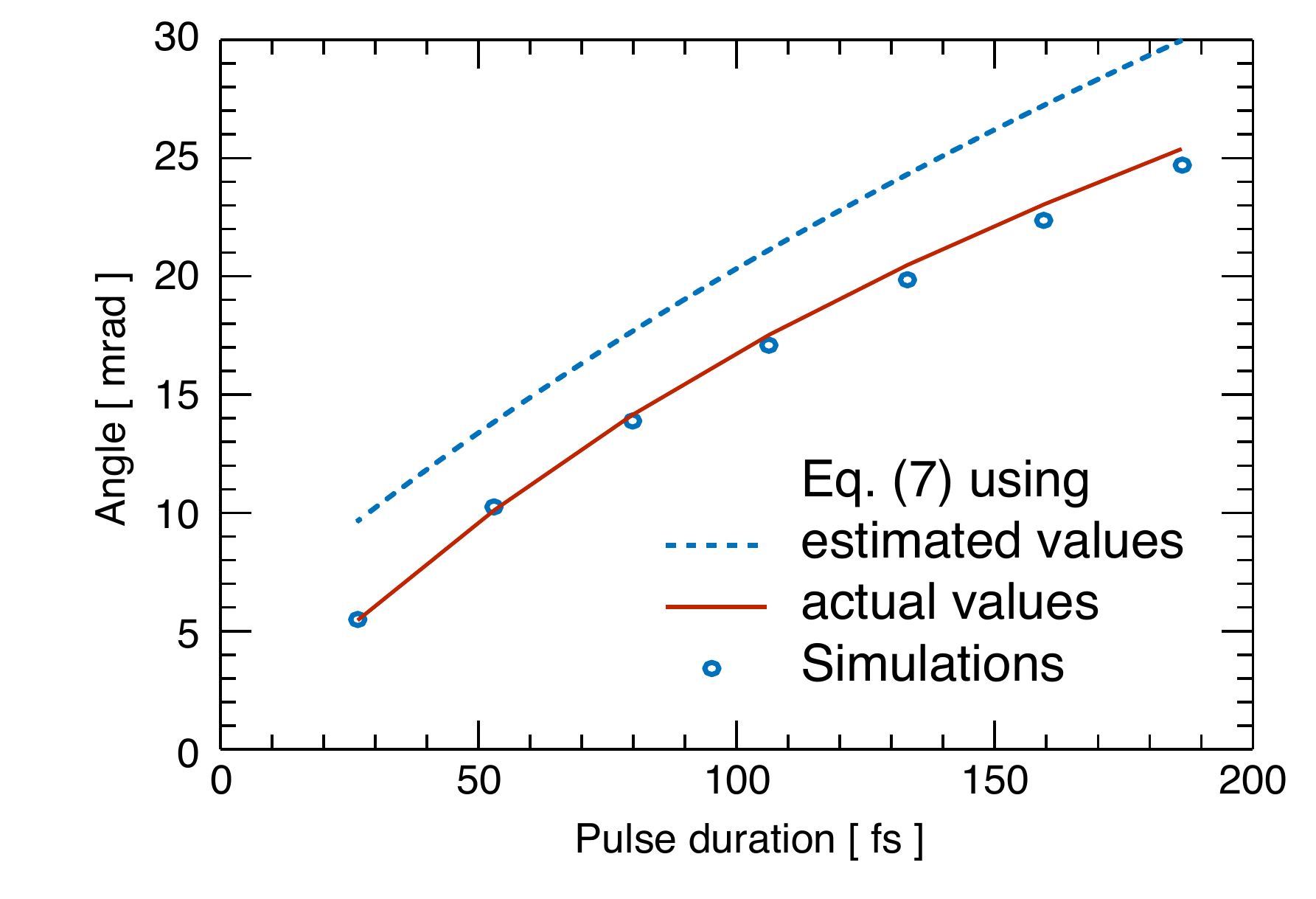}
	\caption{Electron beam divergence after the shutdown of the interaction with the laser given by Eq. (\ref{divergence}) and from simulations.}
	\label{fig_divergence}
\end{figure}

Note that the electron beam energy bandwidth is bound to rise due to stochasticity in the quantum regime $\chi_e \gtrsim 1$ \cite{Ridgers_quantumRRnew, Piazza_qed_energyspread}.
But, the high-energy electrons, on average, radiate more than low-energy electrons. 
This tends to reduce the energy bandwidth, even in the quantum regime of interaction. 
The electron distribution function either spreads, or shrinks, depending on the local conditions.
In the limit where the scattering is still Thompson in the electron rest frame, one can derive an expression for an instantaneous "turning point" \cite{Vranic_quantumRR}.
If the standard deviation of the electron energy distribution function $\sigma$ is larger than $\sigma_T$, then the electron distribution function shrinks. 
For $\sigma<\sigma_T$, stochasticity dominates and the bandwidth of the electron energy distribution rises. 
The value of $\sigma_T$ is given by  \cite{Vranic_quantumRR}
\begin{equation}\label{turning_point_eq}
\sigma_T [mc^2] \simeq 1.4\times10^{-2}~\gamma_T^{3/2}~ I_{22}^{1/4}
\end{equation}
where $\gamma_T$ is the average value of the instantaneous electron Lorentz factor. 
The validity of Eqs.  (\ref{efinal}) and (\ref{turning_point_eq}) can be extended to the regime where $\chi_e\approx1$ by adding a correction for the electron Gaunt factor \cite{Ridgers_RR_spinetto, niel2017quantum}. 
However, the final expression then becomes more complex, and our aim here is to keep the scaling laws as simple as possible. 
The simplicity allows to estimate an asymptotic energy spread \cite{Vranic_quantumRR} as a function of the initial electron energy $\gamma_0$ and the laser intensity and duration:
\begin{equation}\label{explicit_final_sigma}
\sigma_F [mc^2] \lesssim \left( \frac{1.5\times10^{-4} I_{22}^{1/2} ~\gamma_0^3}{\left( 1+6.1\times 10^{-5}\gamma_0~ I_{22}~ \tau_0[\mathrm{fs}] \right)^3}\right)^{1/2}.
\end{equation} 
Equation (\ref{explicit_final_sigma}) is useful for planning experiments, because it allows for a quick estimate of the expected final width of the electron energy distribution function. 
One can also predict the final divergence of the beam. 
Let us define the divergence as the average deflection angle $\theta_F$ from the main axis of beam propagation. 
In this case, one can estimate this value as
\begin{equation}\label{divergence}
	\theta_F\simeq \sqrt{\frac{2}{\pi}} ~ \frac{a_0}{\gamma_F^2} ~ \sigma_{F}
\end{equation}
Comparisons of Eq. (\ref{divergence}) with QED-PIC simulations are given in Fig.  \ref{fig_divergence}. Blue dashed line represents the $\theta_F$ obtained using values of $\gamma_F$ and $\sigma_F$ Eqs. (\ref{efinal}) and (\ref{explicit_final_sigma}). Red line represents the values obtained with Eq. (\ref{divergence}), but using $\sigma_F$ and $\gamma_F$ measured in the simulation. The electron beam initial energy was 0.85 GeV, and it interacted with a circularly polarized laser of $a_0=27$. All other simulation parameters are given in the Appendix.


\section{Energy conversion to high-frequency light in a two-laser standing wave (linear polarization)}\label{sec:cascade}

The number of photons of a plane wave with the normalized vector potential $a_0$ in a volume of $\lambda_0^3$ can be estimated as
\begin{equation}\label{nphotons_vol}
N_{ph} = \frac{E^2}{4\pi}~\lambda_0^3~\frac{1}{\hbar \omega_0}=\frac{a_0^2 ~a_S^2~ 2\pi^2}{\alpha}\simeq3\times 10^{14} a_0^2.
\end{equation}

Here $a_S=mc^2/\hbar \omega_0=4.12\times 10^5~\lambda_0[\mu \mathrm{m}]$ represents the Schwinger critical field in units normalized to the laser frequency. 
In other words, this defines the dimensionless normalized vector potential that corresponds to a Schwinger field at a given wave frequency. 
For our calculation, the total number of available photons in a volume of $\lambda_0^3$ given by Eq. (\ref{nphotons_vol}) should be multiplied by a factor of 2, as the standing wave is formed by two counter-propagating traveling waves. 

The temporal structure of the standing wave causes a periodic emission with a period of $T_L$. 
The constructive interference for magnetic and electric field is temporally spaced at $T_L/4$. 
Periods of particle acceleration (when E is large) and rotation (when B dominates) therefore alternate every $T_L/4$.
The characteristic cycle of emission is $T_L/2$ with the second half of the laser period repeating exactly the same particle dynamics in the opposite direction of motion. 

\begin{figure}
	\includegraphics[width=0.45\textwidth]{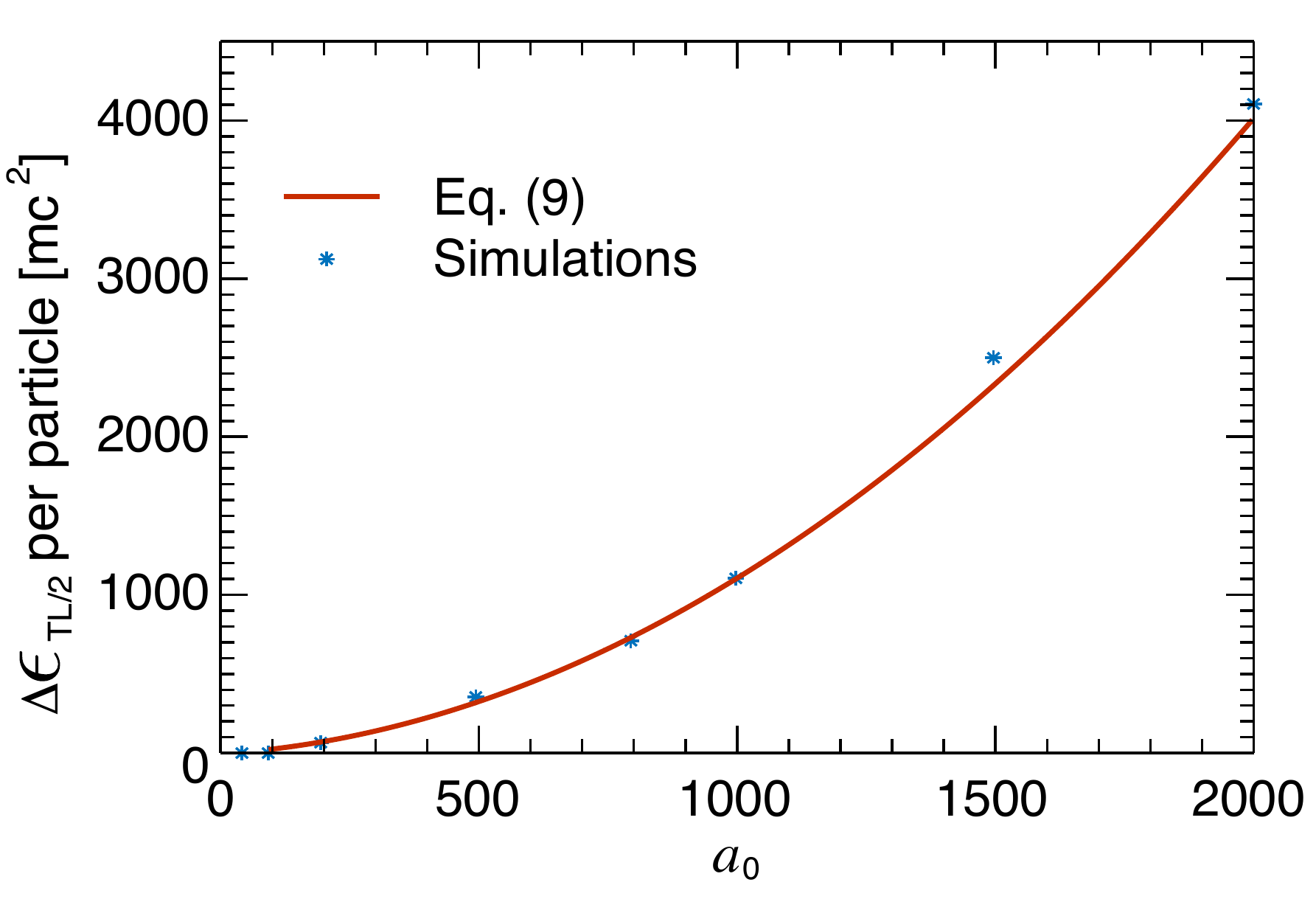}
	\caption{Energy emitted per particle in an undisturbed standing wave during half of a laser period $T_L/2$. The line is given by Eq. (\ref{one_part_half_period}), and points are obtained from simulations.}
	\label{fig_one_part_abs}
\end{figure}

\begin{figure}
	\includegraphics[width=0.45\textwidth]{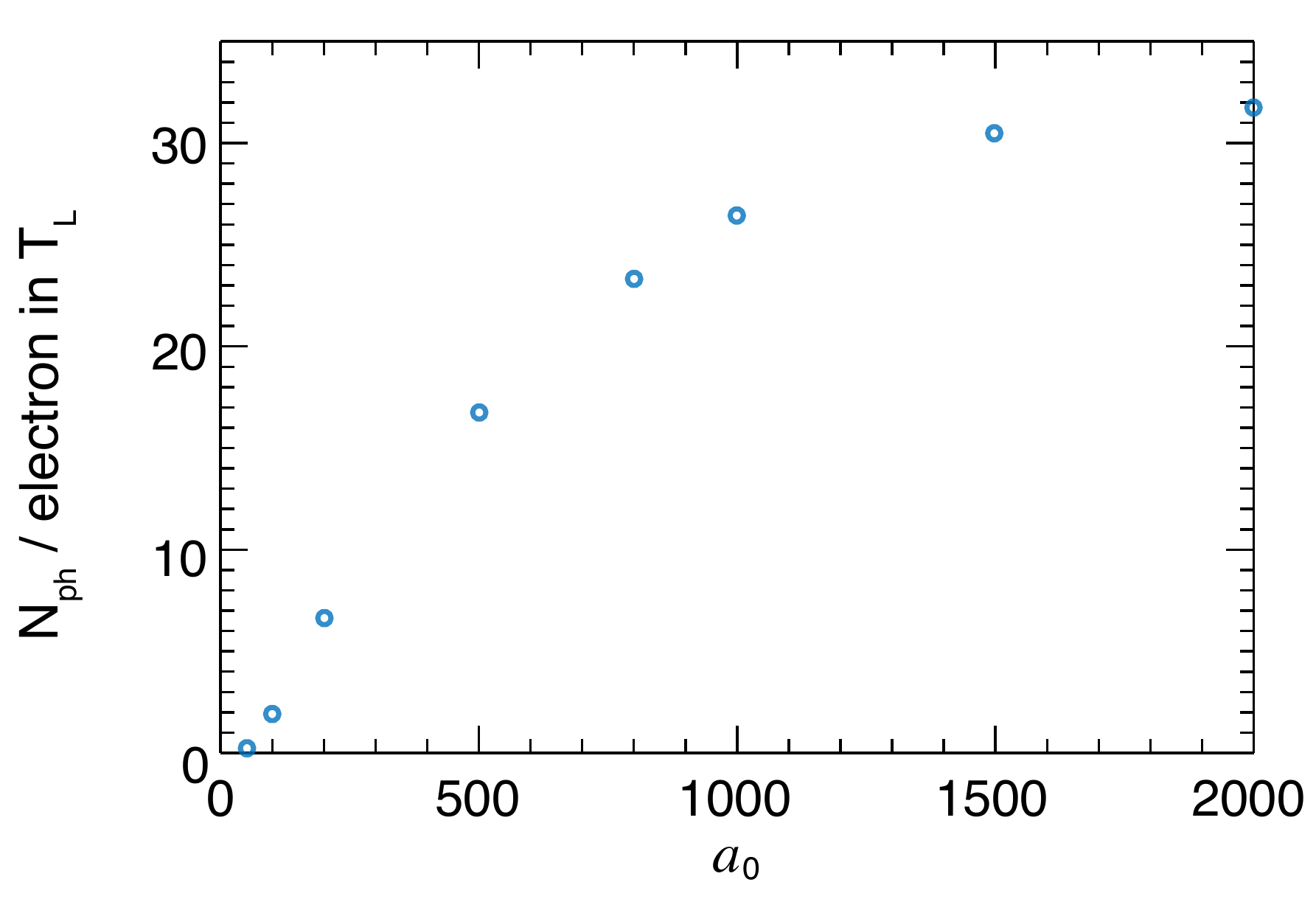}
	\caption{Number of photons with energy above 1 MeV emitted during one laser period $T_L$ per particle (electron or positron). The numbers are given for undisturbed standing wave.}
	\label{fig_nhardphot}
\end{figure}

Particle trajectories are chaotic \cite{Esirkepov, Mendoca1983} and stochastic emission does not allow for a general analytical estimate of the radiated energy using the trajectories alone.
However, average energy absorbed per particle  during the half-period $T_L/2$ can be approximated using ideal simulations where pair production and current deposition are suppressed (see Figs. \ref{fig_one_part_abs} and \ref{fig_nhardphot}).
In such simulations, the wave is not disturbed by the presence of the plasma.
Photons do not decay into pairs, but the particles do experience quantum radiation reaction due to the hard photon emission.

The energy emitted by one particle during $T_L/2$ in an undisturbed linearly polarized standing wave with $\lambda_0=1~\mu$m can be approximated as
\begin{equation}\label{one_part_half_period}
\Delta\gamma \approx 9\times 10^{-4}a_0^2+0.2 a_0. 
\end{equation}
Equation (\ref{one_part_half_period}), obtained as a best fit to the available data, can then be used to estimate the importance of radiative laser absorption for non-ideal scenarios. 
What we mean by radiative absorption is the energy that was transferred from the laser to the particles through classical acceleration, and then radiated to hard photons. 
When the standing wave is interacting with a plasma target, this is one of the possible depletion channels. 

Let us assume that our target is $1~\mu$m thick and the density $n$ is much lower than the relativistic critical density $a_0 n_c$. 
In that case, we do not expect the standing wave to be disturbed. 
Also, by taking a narrow plasma slab, all plasma electrons (and eventually positrons) are located within one laser wavelength.
This limits the interaction of each section of the traveling waves with the plasma to one full laser period. 
The radiative laser depletion can therefore be estimated locally in a $\lambda_0^3$ volume and during one full laser cycle. 
The percentage of the wave turned into high-frequency radiation depends on the local plasma density and the local laser intensity. 
The density gives an estimate of the total number of particles in the $\lambda_0^3$ volume, while the emitted energy per particle depends on the laser intensity and is given by  Eq. (\ref{one_part_half_period}).

If $n=n_c$ and we consider an electron-positron plasma, we have about $10^9$ electrons and $10^9$ positrons  in the  $\lambda_0^3$ volume. The fraction of the laser energy absorbed locally during one laser period is then given by 
\begin{equation}\label{abs_percentage_first}
\left(\frac{\Delta\epsilon}{\epsilon}\right)_{T_L} \approx 3.3~ a_S\times 10^{-6}\left(\frac{n}{n_c}\right)\frac{ (9\times 10^{-4}a_0+0.2)}{a_0}\times 4.
\end{equation}

Here, multiplication factor 4 comes from moving from half of $T_L$ to a full $T_L$ and considering two species, electrons and positrons. 
If we now follow any point of the individual traveling waves at the speed of light as it passes the interaction region, we note that any absorption that happens to that portion of the wave happens within one $T_L$. 
The rest of the time, the wave propagates freely. 
Every $T_L$, there is a fresh portion of the wave interacting with the plasma, while the previous has escaped the interaction region. 
This means that the ratio given by Eq. (\ref{abs_percentage_first}) is approximately equal to the global absorption of the wave after the shutdown of the interaction. 

\begin{figure}
	\includegraphics[width=0.45\textwidth]{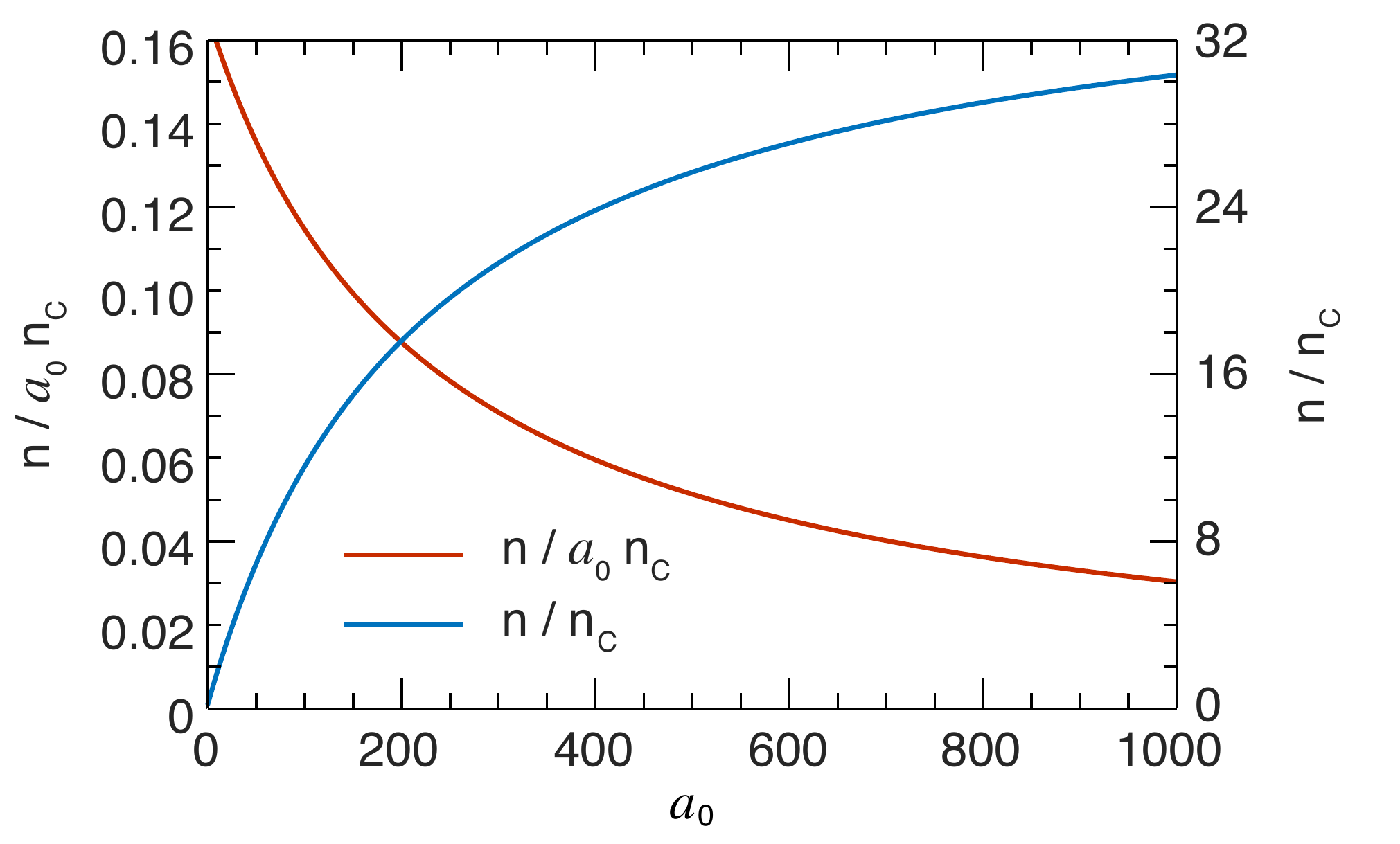}
	\caption{Target density for strong absorption as a function of laser intensity. The target density is calculated when Eq. (\ref{abs_ratio}) predicts 10\% absorption. For controlled conditions, one should aim at least one order of magnitude lower initial target density.  }
	\label{fig_strong_abs}
\end{figure}

For optical lasers with $\lambda_0=1~\mu$m, interacting with a pair plasma we get that in an undisturbed standing wave, the energy converted to high frequency radiation can be approximated as
\begin{equation}\label{abs_ratio}
\left(\frac{\Delta\epsilon}{\epsilon}\right) \approx \frac{ 3\times(9\times 10^{-4}a_0+0.2)}{a_0} \left(\frac{n}{n_c}\right).
\end{equation}
This estimate is not the whole picture, because we neglect the self-consistent fields in the plasma and pair production. 
However, Eq. (\ref{abs_ratio}) allows to estimate how strong is the high-frequency emission as a depletion channel when considering plasma densities much lower than the relativistic critical density $a_0 n_c$. 
Figure \ref{fig_strong_abs} illustrates the strong radiative absorption limit as a function of intensity as predicted by Eq. (\ref{abs_ratio}). 
The characteristic density is calculated for each laser intensity assuming 10\% laser absorption and solving for $n$. 
If we would like to stay in the regime of low absorption (undisturbed wave), we have to use the density below this limit. 
It is interesting to note that one does not need to reach the relativistic critical density for extreme absorption according to this. 
For $a_0=1000$ even $n=160n_c$ is enough for strong absorption.
Cryogenic targets available in experiments have a density on the order of $10~n_c$, which means it is possible to use them for controlled radiation sources, provided that the QED cascade does not increase the plasma density by an order of magnitude. 

\begin{figure}
	\includegraphics[width=0.45\textwidth]{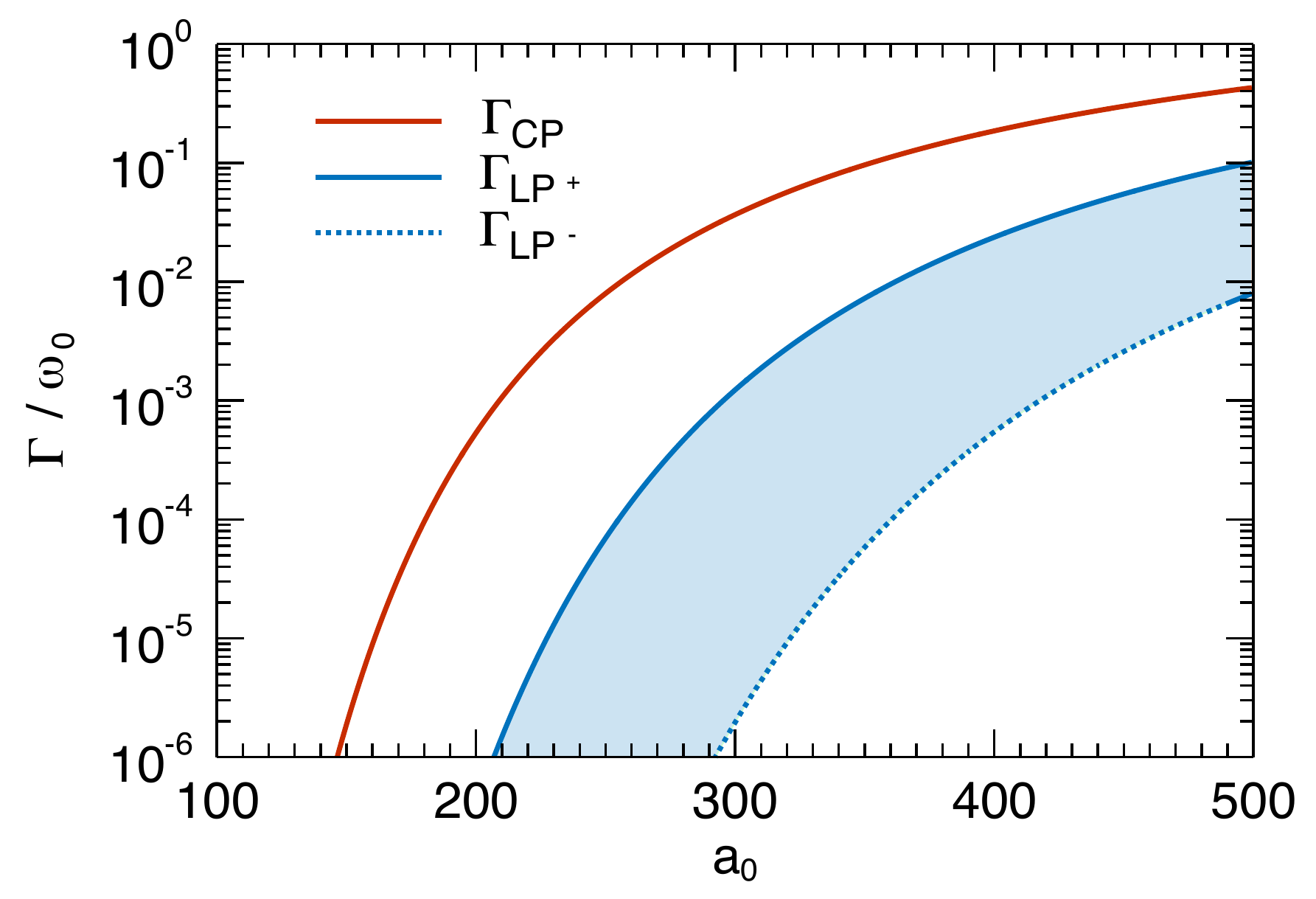}
	\caption{Pair production growth rates for circular and linear polarization given by Eqs. (\ref{growth_CP}) - (\ref{growth_LP_m}).}
	\label{fig_growth}
\end{figure}

To evaluate a parameter range where this is the case, we should analyze also the pair production. 
Plasma density in a cascade increases exponentially: $n=n_0~\exp (\Gamma t)$, where $\Gamma$ is a growth rate that depends on the laser intensity. 
For $a_0<500$, and circularly polarized lasers, the growth rate has an approximate expression given by \cite{ThomasQED}
\begin{equation}\label{growth_CP}
\frac{\Gamma_{CP}}{\omega_0}=2.6\times 10^{-3} a_0 \exp\left(-\frac{2 a_S}{3 a_0^2}\right).
\end{equation}
The growth rate predicted by Eq. (\ref{growth_CP}) is in agreement with the recent calculations by Kostyukov et al\cite{kostyukov_nerush_2018growth} for $a_0\lesssim400$. 
An upper estimate for linear polarization can be estimated by assuming the same total energy yields the same growth rate. This is equivalent to using $a_0/\sqrt{2}$ instead of $a_0$ in Eq. (\ref{growth_CP})
\begin{equation}\label{growth_LP_p}
\frac{\Gamma_{LP+}}{\omega_0}=1.8\times 10^{-3} a_0 \exp\left(-\frac{4 a_S}{3 a_0^2}\right).
\end{equation}
Another option for Gaussian laser pulses is to insert $a_0/2$ instead of $a_0$ in Eq. (\ref{growth_CP}) to account for the fact that not all particles experience the instantaneous maximum intensity
\begin{equation}\label{growth_LP_m}
\frac{\Gamma_{LP-}}{\omega_0}=1.3\times 10^{-3} a_0 \exp\left(-\frac{8 a_S}{3 a_0^2}\right).
\end{equation}
Using $a_0/2$ is also more consistent with the data obtained from cascade simulations at higher intensities  \cite{ThomasQED,Thomas_POP_2016} ($a_0>1000$). 
Equations (\ref{growth_LP_p}) and (\ref{growth_LP_m}) define a range of values for expected growth rate at each $a_0$ for linear polarization. 
We should note here that near-threshold pair production is very sensitive to the seeding \cite{Jirka, ThomasQED, tamburini}, so one may not be always able to reach the given growth rates. 
When using laser pulses, the actual growth rate is likely to be closer to $\Gamma_{LP-}$ than to $\Gamma_{LP+}$. 
For example, the multiplicity we get for a 24.5 fs laser pulse with peak intensity of $a_0=500$ using Eq. (\ref{growth_LP_m}) is 1.44, while the simulation data give 1.69. 
For the same data, the upper bound $\Gamma_{LP+}$ predicts a multiplicity of $\sim$ 100, which is two orders of magnitude higher.
Using an infinite plane wave, the multiplicity in the simulation increases, but it stays on the same order as predicted by $\Gamma_{LP-}$ in Eq. (\ref{growth_LP_m}). 
The growth rate estimates for $a_0<500$ given by Eqs. (\ref{growth_CP})-(\ref{growth_LP_m}) are illustrated in Fig \ref{fig_growth}. 

As we mentioned before, for designing a radiation source, it may be favourable to keep the pair production rate low and operate in a more controlled emission regime. 
To avoid the uncertainty of seeding, it is also important to have a reliable estimate regarding how many particles are contained within the interaction volume. 
In this sense, using a gas jet would not be optimal, even though it has an advantage of offering a possibility to start at a low density. 
A dense solid target can be manufactured to high precision, but then an undisturbed standing wave interaction is not possible.
The best option available to date could be to use cryogenic targets, composed of hydrogen ice.
They have recently been used in ion acceleration experiments, and $\mu$m-level thickness was achieved \cite{Lotti_cryogenic, Cryogenic_freestanding} at a density $\sim 10~n_c$. 

In the following paragraphs, we consider the output radiation generated by interaction of two pulses with one such target, initially composed of electrons and protons.  
For completeness, we compare the findings also with the case when the  target is initially composed of electrons and positrons, as well as the case when the standing wave is undisturbed. 
A range of intensities between $a_0=100$ and $a_0=1500$ was considered.

\begin{figure*}
	\includegraphics[width=1.0\textwidth]{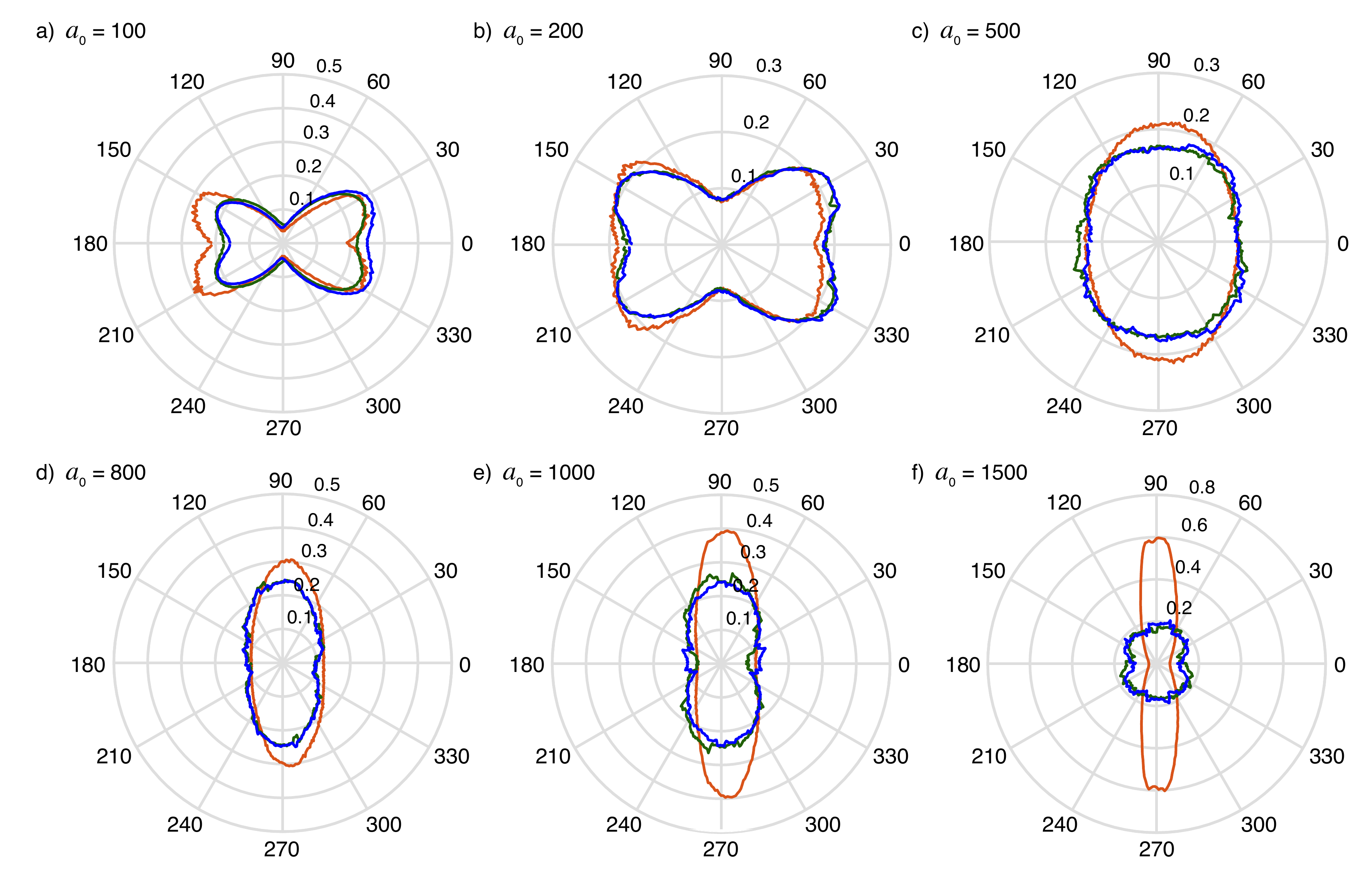}
	\caption{Angular distribution of radiation as a function of intensity. The values of the radius represent a fraction of total energy radiated in a given direction per 1 rad (i. e. isotropic radiation would correspond to a circle with a radius of $1/2\pi$). The red curves correspond to the radiation in an undisturbed standing wave. The other two curves correspond to the radiation pattern obtained using a 10 $n_c$ target composed of (green) electrons and positrons or (blue) electrons and protons. }
	\label{fig_polar}
\end{figure*}

\begin{figure*}
	\includegraphics[width=0.5\textwidth]{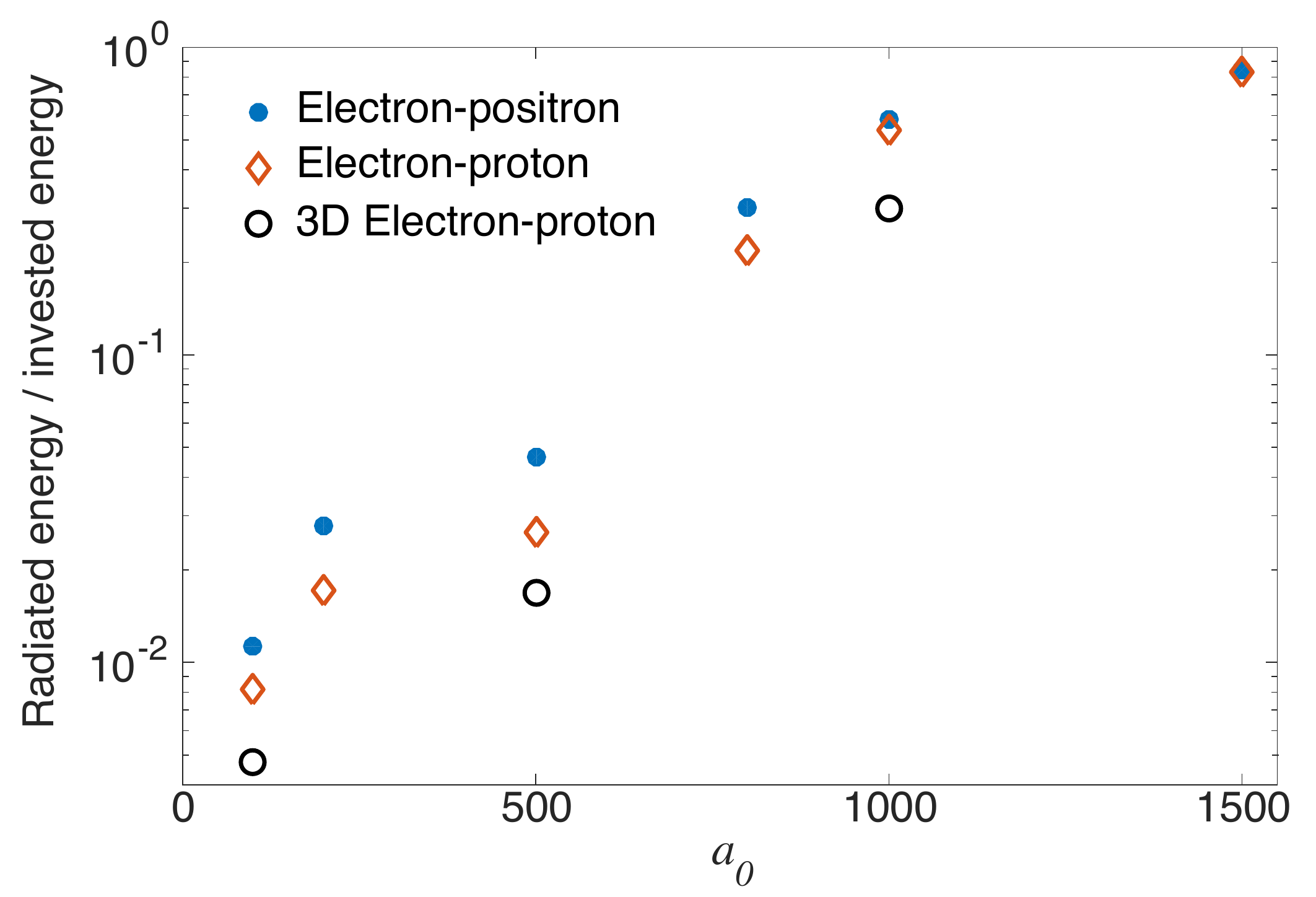}
	\caption{Conversion efficiency of laser energy to emitted radiation using a 1 $\mu$m-thick target with an initial density of 10 $n_c$. The results shown are for a target composed of electrons and positrons (blue dots) or electrons and protons (red diamonds).}
	\label{fig_abs_10nc}
\end{figure*}

Figure \ref{fig_polar} displays the angular distribution of radiation for all the cases, while Fig. \ref{fig_abs_10nc} shows the conversion efficiency of optical light to high-frequency radiation. 
The differences between the electron-proton and electron-positron target in the typical radiation directions are almost negligible, although there is a difference in the energy conversion efficiency. 
This is not surprising, because in the electron-positron target, there are twice as many radiating leptons. 
The angular distribution in an undisturbed standing wave is quite different at some intensities compared to that of the 10 $n_c$ target. 
For $a_0=100$, the 10 $n_c$ target represents already 10\% of the relativistic critical density for such a wave. 
Some level of discrepancy is therefore expected for the lower end of the explored intensities. 
At higher intensities, the discrepancies come from the pair production that increases the target density during the interaction, until the density is high enough for wave disruption. 
Mid-range intensity $a_0\sim500$ seems to be the best choice for controlled emission, because the pair multiplicity is low, and at the same time the intensity is high enough not to be too disturbed by the presence of the target. 

\begin{figure*}
	\includegraphics[width=1.0\textwidth]{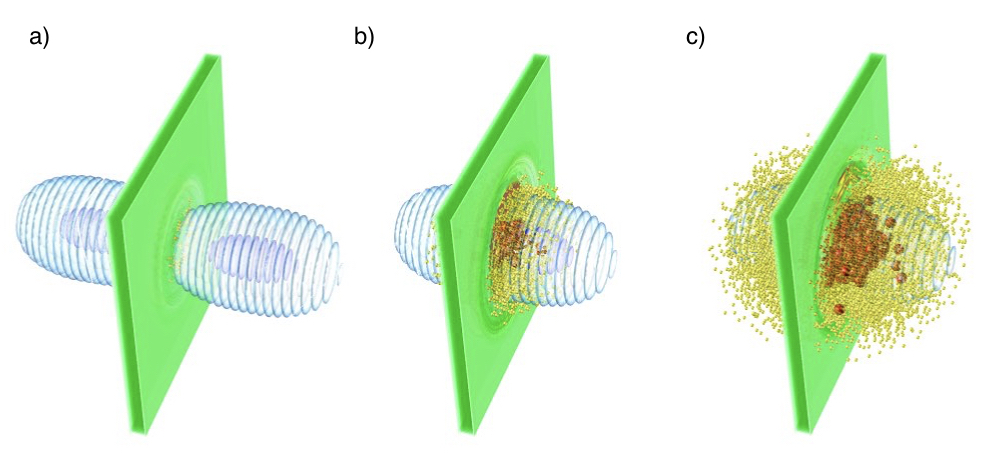}
	\caption{3D simulation of a two-laser cascade produced using a cryogenic ice target and two lasers of $a_0=1000$. Half of the total laser energy is converted to $\gamma$-rays.}
	\label{fig_3d}
\end{figure*}

Figure \ref{fig_3d} shows a 3D simulation using a cryogenic target for $a_0=1000$. 
The conversion efficiency obtained in 3D simulations for $a_0=100$, $a_0=500$ and $a_0=1000$ are displayed together with 2D results in Fig. \ref{fig_abs_10nc}. The absorption is somewhat lower in 3D (as the average laser intensity is lower), but is of the same order of magnitude, so the conclusions regarding the different regimes of interaction are consistent with these results as well.

\section{Conclusion}

We have shown that it is possible to convert most of the energy from the laser (or interacting electrons in a laser-electron scattering) to $\gamma$-rays in the near-future laser experiments. 
To achieve a strong conversion efficiency, one should either use a solid-density thin target, or use intensities that can initiate a QED cascade that produces enough pairs to increase the plasma density during the interaction ($a_0\sim1000$).
It is also possible to obtain radiation emission in controlled conditions. 
Using cryogenic targets, this is possible for $a_0\sim500$, where the number of particles is not expected to increase more than twice due to the pair production, and the target density $n\sim 10 n_c$ is small enough to keep the radiative absorption below 10\%.

\section{Appendix}
All simulations are performed with the QED module of OSIRIS \cite{OSIRIS}.
The QED module is a Monte-Carlo module that accounts for the photon emission and Breit-Wheeler pair production, and is implemented as an addition to the standard PIC loop of OSIRIS. Energetic photons are initialized as an additional particle species. 
The emission rates are found in Refs \cite{pair_rate1,pair_rate2,pair_rate3}. OSIRIS QED has been used previously in Refs. \cite{ThomasQED, Thomas_POP_2016, MV_ppcf2016, Vranic_quantumRR,vranic2018_scirep}. A similar method for incorporating BW pair production is used in several other codes  \cite{Elkina_rot, Ridgers_solid, Nerush_laserlimit, Bell_Kirk_MC, Gonoskov2015review, Gonoskov2014, Gremillet, lobet,  Basmakhov}.

The simulations from Section \ref{sec:dev_cas} displayed in Fig. \ref{divergence}  are performed with parameters as in Vranic et al\cite{Vranic_quantumRR}. The electron beam initial energy is 0.85 GeV, and initial beam divergence is $p_\perp/p_\parallel\sim 0.2$ mrad. The laser is transversely a plane wave with a temporal envelope. The total pulse duration is given by $\tau=\tau_\mathrm{flat}+(\tau_\mathrm{rise}+\tau_\mathrm{fall})/2$, where $\tau_\mathrm{flat}$ is the constant amplitude section of the wave that was varied from 0 to 160 fs. The envelope function has a smooth rise and fall, the same for all the simulations: $\tau_\mathrm{rise}=\tau_\mathrm{fall}=26.6$ fs. The simulations are performed in 2D, with a box size of $500\times20~ c^2/\omega_0^2$ resolved with $5000\times200$ cells and the timestep $dt=0.04~\omega_0^{-1}$ using 16 particles per cell (ppc). 

Section \ref{sec:cascade} has several types of simulations. Ideal simulations displayed in Figs. \ref{fig_one_part_abs}, \ref{fig_nhardphot} and \ref{fig_polar} were performed with no current deposition and pair production: the standing wave is undisturbed by the presence of the plasma and the plasma density does not grow due to the new particle generation. 
The transverse boundary conditions are periodic. 
The simulation box size is $200\times10~c^2/\omega_0^2$, resolved with $2000\times100$ cells and 9 ppc. 
The plasma slab is $1~\mu$m thick, and is composed of electrons and positrons. The initial density is $n=0.001~n_c$.  
The two lasers have a 2-cycle smooth rise and fall, and a 10-cycle flat section. 
All the measurements in ideal conditions were taken while all particles are fully immersed in the flat-top section of the standing wave.

The 2D and 3D simulations shown in Figs. \ref{fig_polar}, \ref{fig_abs_10nc}, and \ref{fig_3d} are performed including all options of the QED PIC module. In 2D (3D) the box size was $200\times192~c^2/\omega_0^2$ ($200\times192\times192~c^3/\omega_0^3$), resolved with $2000\times1920$ ($2000\times960\times960$) cells and 9 ppc (27 ppc). 
The timestep is $dt=0.005~\omega_0^{-1}$ and boundary conditions are open in all directions.
The two laser pulses have a Gaussian transverse profile, with a spotsize of $W_0=3~\mu$m. 
The temporal envelope slope is defined by a polynomial function $f(t)=10(t/\tau_0)^3-15(t/\tau_0)^4+6(t/\tau_0)^5$, where the pulse duration is $\tau_0=25$ fs.
The plasma slab is initially 1 $\mu$m wide, with density of $n=10~n_c$ and composed of either electrons and positrons, or electrons and protons.

\begin{acknowledgments}
 This work is supported by the European Research Council (ERC-2015-AdG Grant 695088), Portuguese Science Foundation (FCT) Grant SFRH/BPD/119642/2016 and German Research Foundation (DFG) Grant Nr. 361969338. We acknowledge PRACE for awarding access to MareNostrum based in Barcelona Supercomputing Centre. Simulations were performed at the IST cluster (Lisbon, Portugal), and MareNostrum (Spain).
 \end{acknowledgments}


%


\end{document}